\documentclass[reprint,eqsecnum,floats,aps,amsmath,amssymb,nofootinbib,prd,onecolumn, showpacs]{revtex4-1}
\usepackage{eufrak}
\usepackage{graphicx}
\usepackage{amsmath,amssymb}
\usepackage{hyperref}
\usepackage{graphicx}
\usepackage[utf8x]{inputenc}

\begin{document}
	
\author{Beatriz Elizaga Navascu\'es}
\email{beatriz.b.elizaga@gravity.fau.de}
\affiliation{Institute for Quantum Gravity, Friedrich-Alexander University Erlangen-N{\"u}rnberg, Staudstra{\ss}e 7, 91058 Erlangen, Germany}
\author{Guillermo A. Mena Marug\'an}
\email{mena@iem.cfmac.csic.es}
\affiliation{Instituto de Estructura de la Materia, IEM-CSIC, Serrano 121, 28006 Madrid, Spain}
\author{Santiago Prado}
\email{santiago.prado@iem.cfmac.csic.es}
\affiliation{Instituto de Estructura de la Materia, IEM-CSIC, Serrano 121, 28006 Madrid, Spain}

\title{Fock quantization of the Dirac field in hybrid quantum cosmology: Relation with adiabatic states} 

\begin{abstract} 

We study the relation between the Fock representations for a Dirac field given by the adiabatic scheme and the unique family of vacua with a unitarily implementable quantum evolution that is employed in hybrid quantum cosmology. This is done in the context of a perturbed flat cosmology that, in addition, is minimally coupled to fermionic perturbations. In our description, we use a canonical formulation for the entire system, formed by the underlying cosmological spacetime and all its perturbations. After introducing an adiabatic scheme that was originally developed in the context of quantum field theory in fixed cosmological backgrounds, we find that all adiabatic states belong to the unitary equivalence class of Fock representations that allow a unitarily implementable quantum evolution. In particular, this unitarity of the dynamics ensures that the vacua defined with adiabatic initial conditions at different times are unitarily equivalent. We also find that, for all adiabatic orders other than zero, these initial conditions allow the definition of annihilation and creation operators for the Dirac field that lead to some finite backreaction in the quantum Hamiltonian constraint and to a fermionic Hamiltonian operator that is properly defined in the span of the \textit{n}-particle/antiparticle states, in the context of hybrid quantum cosmology. 

\end{abstract}

\pacs{04.62.+v, 98.80.Qc,  04.60.Pp.}

\maketitle 

\newpage

\section{Introduction}

There exists an inherent difficulty to select a vacuum state with acceptable physical properties for fields that propagate in generic curved spacetimes, even when one uses well known Fock representations in their quantization \cite{Wald}. This ambiguity is closely related to the fact that the notion of particle, as one defines it in quantum field theory (QFT), is nebulous even in the presence of a large number of symmetries. This problem is often overlooked in standard QFT in Minkowski spacetime because the Poincaré vacuum plays then a privileged role, directly tied up to the observation that flat spacetime is maximally symmetric \cite{Peskin}.  In this sense, a central question in any scheme pursuing the Fock quantization of matter fields in a generic spacetime background is the specification of the physical properties that the corresponding vacuum must possess. This issue has been studied at great length for free scalar linear fields \cite{Wald,bidav},  but much less for fermionic fields, such as the Dirac field \cite{Parker}. 

For cosmological spacetimes, a traditional line of attack to the problem of the choice of vacuum is the adiabatic proposal \cite{adiabatic1,adiabatic2,adiabatic3}, which in recent times has found formal support in the algebraic approach to QFT \cite{haag}. In this approach, one chooses a series of observables and specifies the relations among them, something which includes the dynamics and the standard commutation (or anticommutation for fermionic fields) relations, in such a way that one constructs a *-algebra. A state is then a normalized positive linear functional from this *-algebra to the complex numbers, which can be interpreted as the result of taking the expectation value of the observables on a physical state. In many cases, a specific Fock representation can be recovered from each algebraic state by means of the so-called GNS construction  \cite{gn,se}. A set of states that is traditionally favored in this approach is formed by the Hadamard states, which are characterized by a very specific singularity structure of their two-point function  \cite{Hadamard,Hadamard2}. In particular, their associated energy-momentum tensor has good renormalizability properties. The adiabatic scheme aims to provide a strategy to approximate Hadamard states in cosmology by solving the differential equations of motion of the field in an iterative way, with the hope that, if the iteration converges, one would obtain in the end a true Hadamard state. Actually, for scalar fields propagating in standard Friedmann-Lema\^{\i}tre-Robertson-Walker (FLRW) cosmologies it turns out that all adiabatic states are locally quasiequivalent to a Hadamard state \cite{luders,junker}. The complications that arise in this scheme are well known in the case of scalar fields in cosmological backgrounds, as the iterative relations may not converge for general cosmological evolutions. For Dirac fields in cosmological spacetimes, a similar level of consensus on the definition of adiabatic states and their properties has not been reached \cite{landete,barbero,hollands,najmi}. 

Over the last decade, an alternative strategy has been put forward in order to reduce the ambiguities in the choice of a vacuum for fields in cosmological spacetimes \cite{uni1,uni2}. In addition to symmetry considerations, this strategy rests primarily on the criterion that the annihilation and creation operators of the Fock quantization display an evolution that is unitarily implementable. This criterion has been shown to select a unique family of vacua, related to each other by unitary transformations, on a multitude of cosmological scenarios \cite{uni3,uniqueness1,uniqueness2,uniqueness3}, including the case of Dirac fields in a flat FLRW cosmology \cite{uniqfermi}. Actually, this criterion is, in turn, motivated in the context of quantum cosmology by the so-called hybrid approach to the quantization of inhomogeneous systems  \cite{hlqc1,hlqc2,hlqc2}, which is based on a splitting of the phase space into a homogeneous and an inhomogeneous sector, in a way that is specially suitable to obtain a well-behaved dynamics for the complete cosmology. Then, one quantizes the inhomogeneous degrees of freedom employing a Fock representation with nice ultraviolet properties, and the homogeneous geometry with techniques inspired by a certain canonical approach to quantum cosmology (for instance, the formalism known as Loop Quantum Cosmology \cite{LQC}, inspired by Loop Quantum Gravity \cite{Thiemann}). In this context, one can actually restrict the choice of the Fock vacuum even more, exploiting the freedom allowed by the hybrid approach in the way to split the degrees of freedom into the homogeneous and the inhomogenous sectors that are  to be quantized. Indeed, this was first done for fermionic perturbations in inflationary cosmologies \cite{fermihlqc} in an attempt to find a representation such that some kind of quantum backreaction on the homogeneous cosmological sector remains finite without the need of a regularization scheme, and that one gets a Hamiltonian constraint which is properly defined on the dense set of the Fock space spanned by the \textit{n}-particle/antiparticle states \cite{backreaction}. Additionally, it is possible to further refine the description of the inflationary cosmology and arrive at a recurrence relation by which the dynamics of the annihilation and creation operators that describe the fermionic, scalar, and tensor perturbations become diagonal in the asymptotic limit of infinitely large particle/antiparticle wave numbers \cite{diagfermi,diagscalar}. 

This paper aims to bridge the gap between the two schemes commented above for the choice of a Fock vacuum in the case of a Dirac field minimally coupled to a flat FLRW cosmology with compact hypersurfaces. For that, we will adapt the adiabatic scheme for the fermionic field presented in the Dirac representation in Ref. \cite{barbero}, inspired in turn by Ref. \cite{landete}, to the Weyl representation employed so far in hybrid quantum cosmology. We will compare these adiabatic vacua with those of the family of unitarily equivalent Fock representations that arise from the annihilation and creation operators defined in hybrid quantum cosmology, restricted to the context of QFT in curved spacetimes. The fundamental result that we will obtain is that all adiabatic states belong in fact to this equivalence family, and that, for adiabatic orders greater than zero, they allow the definition of annihilation and creation operators in hybrid quantum cosmology that produce finite backreaction terms in the Hamiltonian constraint and give rise to a properly defined Hamiltonian operator. Furthermore, in the context of QFT, the unitary implementability of the dynamics in such Fock quantizations guarantees that the states constructed with adiabatic initial conditions at different times of the cosmological evolution are all unitarily related. Finally, in the Appendix, we will briefly analyze the adiabatic approach proposed by Hollands in Ref. \cite{hollands} from an algebraic perspective, and argue that there generally exist obstructions for its implementation to define Fock vacua.

The structure of this paper is organized as follows. In Sec. \ref{model} we introduce the physical model, which consists of a Dirac field treated as a perturbation around a flat, inflationary FLRW cosmology, and then we summarize the main properties of the choices of annihilation and creation operators for the quantization of this fermionic field in the hybrid approach. In Sec. \ref{adiabatic} we apply the adiabatic scheme to fermions in the Weyl representation. Sec. \ref{unit} is devoted to the comparison of these adiabatic states with those associated with the choices of annihilation and creation operators selected in hybrid quantum cosmology. We show that all the adiabatic states determine Fock representations that are unitarily equivalent to those of the hybrid quantization. We summarize our conclusions in Sec. \ref{conclusions}. The obstructions found in the adiabatic scheme of Ref. \cite{hollands} are discussed in the Appendix. Throughout the paper, we employ units such that $\hbar=c=G=1$. 

\section{Physical system and properties of the quantization}\label{model}
   
Let us start by describing the spatially homogeneous part of our system. We consider a flat FLRW spacetime geometry specified by a scale factor ${\tilde a}$. The spatial sections that foliate this cosmology are compact and isomorphic to the three-dimensional torus $T^3$. As the matter content that fuels the dynamics of this cosmological geometry, we minimally couple a homogeneous scalar field (inflaton) $ \tilde{\phi}$ subject to a potential $V( \tilde{\phi})$. 

In this cosmological model, we include a Dirac field with mass $M$ that is treated entirely as a perturbation (including its homogeneous component, if there is one). In order to obtain a satisfactory Hamiltonian formulation of the entire system, and contemplate the possibility of making canonical transformations that mix the homogeneous and fermionic sectors, we truncate the action at quadratic order in these perturbations \cite{fermihlqc,DH}. One may also include perturbations (of the same magnitude) of the spacetime metric and the inflaton field, describing small anisotropies and inhomogeneities. Nonetheless, we will obviate them in our analysis because, at the considered order of truncation, they do not couple to the fermionic contribution that we want to study. The truncated perturbative action supplies the canonical structure and the constraints needed to construct a Hamiltonian description of the whole system.
 
To work with the Dirac field, we use the Weyl representation of the constant generators $\gamma^{b}$, $b=0,...,3$, of the Clifford algebra associated with the $4$-dimensional Minkowski metric, namely
\begin{align}
\gamma^{0}=i
\begin{pmatrix}
0 & I \\ I & 0
\end{pmatrix},\qquad \vec{\gamma}=i
\begin{pmatrix}
0 & \vec{\sigma} \\ -\vec{\sigma} & 0
\end{pmatrix},
\end{align}
where $I$ is the $2$-dimensional identity matrix,  $\vec{\gamma}=(\gamma^{1},\gamma^{2},\gamma^{3})$, and $\vec{\sigma}=(\sigma_1 ,\sigma_2 ,\sigma_3)$ is the tuple formed by the three Pauli matrices. After imposing the time gauge on the tetrad of the homogeneous cosmology (so that the corresponding triad has no internal time components \cite{fermihlqc}), we rescale the Dirac field by $\tilde{a}^{3/2}$ in order to get constant Dirac brackets between this field and its complex conjugate. In addition, we exploit the symmetries of the homogeneous spatial sections of the cosmological spacetime by expanding each of the two chiral components of the fermionic field in a complete set of eigenespinors of the Dirac operator $-i\vec{\sigma}\vec{\nabla}$ on $T^3$. These eigenspinors can be divided into two subsets according to their helicity, with label $\lambda=\pm 1$. Since the torus is compact, the spectrum of the Dirac operator is discrete, with eigenvalues $\lambda\omega_k$, where $\omega_k= 2\pi\vert \vec k +\vec \tau\vert/l_0$, $\vec k\in \mathbb{Z}^3$, $2\vec\tau$ can be any of the vertices of the unit cube and characterizes the spin structure, and  $l_0$ is the compactification length of the torus. The eigenspinor associated with $\lambda\omega_k$ has the form (adopting a diagonal fiducial coordinate system)
\begin{align}\nonumber
 \xi_{\lambda}(\vec{k})\exp{[i2\pi(\vec k +\vec \tau)\vec{x}/l_0]},
 \end{align}
where $\vec{x}$ are the spatial coordinates on the torus. The bispinor $\xi_{\lambda}(\vec{k})$ is normalized so that $\xi_{\lambda}^\dagger\xi_{\lambda}=1$. The rescaled Dirac field can then be described by its left-handed and right-handed time-dependent coefficients with helicity $\lambda$ in an eigenspinor expansion. These coefficients can be handled as ordered pairs of Grasmann variables, respectively describing the left-handed and right-handed components of the field and, up to a constant factor $l_0^{-3/2}$, we will call them $(x_{\vec k ,\lambda}, y_{\vec k ,\lambda})$. Each of these mode coefficients displays a non-vanishing Dirac bracket only with its complex conjugate, in that case being equal to $-i$.
 
We can then introduce annihilation-like variables $a_{\vec{k},\lambda}$ for particles and creation-like variables $\bar{b}_{\vec{k},\lambda}$ for antiparticles by means of a canonical transformation of the form \cite{uniqfermi}
\begin{equation}\label{eq: u}
 \begin{pmatrix}
 a_{\vec k ,\lambda} \\ 
 \bar{b}_{\vec k ,\lambda}
 \end{pmatrix} 
 =
 \begin{pmatrix}
 f_1^{k, \lambda} & f_2^{k, \lambda} \\ g_1^{k, \lambda} & g_2^{k, \lambda}
 \end{pmatrix}\left[I-\frac{1-\lambda}{2}(I-\sigma_1)\right]\begin{pmatrix}x_{\vec k ,\lambda} \\  y_{\vec k ,\lambda}
 \end{pmatrix}.
\end{equation}
We do not mix different modes of the Dirac operator and only allow mode dependence of the coefficients of the transformation through $\omega_k$, in order to respect the spatial symmetries of the dynamics \cite{uniqfermi,diagfermi}. Besides, we ask that 
\begin{equation}\label{eq:fg}
f_2^{k, \lambda}=e^{iF_2^{k, \lambda}}\sqrt{1-\left\vert f_1^{k, \lambda}\right\vert^2},\qquad g_1^{k, \lambda}=e^{iJ_{k, \lambda}}\bar f_2^{k, \lambda},\qquad g_2^{k, \lambda}=-e^{iJ_{k, \lambda}}\bar f_1^{k, \lambda},
 \end{equation} 
where $J_{k, \lambda},F_2^{k, \lambda}\in\mathbb{R}$, so that each annihilation and creation-like variable only displays a non-vanishing Dirac bracket equal to $-i$ with its complex conjugate variable, giving rise in this way to standard canonical anticommutation relations for annihilation and creation operators. In our notation, the overbar indicates complex conjugation. In general, we allow for linear combinations \eqref{eq: u} that depend on the homogeneous sector, so that $f_l^{k, \lambda}=f_l^{k, \lambda}(\tilde a,\pi_{\tilde a},\tilde \phi,\pi_{\tilde \phi})$, with $l=1,2$ and the symbol $\pi$ (labeled with a subindex) denoting canonical momenta. Following  Ref. \cite{fermihlqc} (see also Ref. \cite{CMM}), we can complete the above transformation of fermionic variables into a canonical transformation for the whole system, including the FLRW cosmology. For this, we must correct the homogeneous variables in order to arrive to a set $( a,\pi_{ a}, \phi,\pi_{ \phi})$ that is canonical with the annihilation and creation-like variables defined in Eq. \eqref{eq: u}. Each of these definitions of fermionic variables can then be understood as the selection of a particular dynamical splitting of the homogeneous and fermionic degrees of freedom in phase space. In fact, each splitting results into a different identification of the fermionic contribution to the zero-mode of the Hamiltonian constraint \cite{backreaction}, the only non-trivial constraint to which the system is subject. This contribution is, in general, not diagonal, by which we mean that it contains interacting terms of the sort of $a_{\vec k ,\lambda} {b}_{\vec k ,\lambda}$. This is especially relevant upon quantization, because a multitude of important features depend on the behavior of the non-diagonal part of the fermionic contribution to the Hamiltonian constraint in the asymptotic limit of infinitely large $\omega_k$. Indeed, choices of canonical annihilation and creation-like variables that result in a decrease of asymptotic order for the coefficients of these interacting terms turn out to display much better physical properties. 
 
The results about the consequences of the selection of variables for the fermionic perturbations proven in previous works \cite{uniqfermi,backreaction,diagfermi} can be summarized as follows:
\begin{itemize}
\item After one chooses a standard convention for particles and antiparticles, the annihilation and creation-like variables undergo an evolution that is unitarily implementable in the context of QFT in a fixed FLRW cosmology if and only if, in the asymptotic limit of large $\omega_k$ \cite{uniqfermi}:
 \begin{equation}\label{eq: cond1}
f_1^{k,\lambda}=\frac{Ma}{2\omega_k}e^{iF_2^{k,\lambda}}+\theta^{k,\lambda},\qquad \sum_{\vec k\in {\mathbb{Z}}^3}\left\vert\theta^{k,\lambda}\right\vert^2<\infty.
\end{equation}  
This condition ensures that the interacting fermionic part of the Hamiltonian has asymptotic order $\mathcal{O}(\omega_k^{-1})$ \cite{fermihlqc}. Furthermore, all possible families of annihilation and creation operators defined by means of coefficients of the form \eqref{eq: cond1} define unitarily equivalent Fock representations \cite{uniqfermi}.
\item With a hybrid quantization of the entire system, it is possible to identify a quantity, interpretable as a backreaction, that appears in the quantum dynamical equation of the fermionic states and that measures the average difference between the quantum evolution of the perturbed and unperturbed cosmology. Unfortunately, this quantity generally fails to be finite. In this case, rather than regularize by performing a ``substraction of infinities'', one can further restrict the choice of fermionic variables (and therefore the way to split the degrees of freedom in phase space)  so that, asymptotically \cite{backreaction},
\begin{equation}\label{eq: cond2}
\theta^{k,\lambda}=-i\frac{\pi M\pi_a}{3l_0^3\omega_k^2}e^{iF_2^{k,\lambda}}+ \vartheta^{k,\lambda}, \qquad \sum_{\vec k\in {\mathbb{Z}}^3}\omega_k\left\vert\vartheta^{k,\lambda}\right\vert^2<\infty .
\end{equation}
\item One can go one step beyond and demand that the interacting fermionic part of the Hamiltonian be square summable. This happens to be the necessary and sufficient condition for the Hamiltonian constraint to be properly defined in the dense set spanned by the $n$-particle/antiparticle states within Fock space, and amounts to require that the following sequence be summable as well \cite{backreaction}:
\begin{equation}
\{\omega_k^2\left\vert\vartheta^{k, \lambda}\right\vert^2\}_{\vec k \in {\mathbb{Z}}^3}.
\end{equation}
\item The last step in this improvement of the properties of the fermionic Hamiltonian upon quantization is a recursive procedure to diminish, as much as desired, the asymptotic order of its interacting part \cite{diagfermi}. This method of ``asymptotic diagonalization" restricts almost completely the choice of fermionic canonical variables in the ultraviolet regime, leaving all the possible remaining freedom in the determination of the phases  $J_{k,\lambda}$ and $F_2^{k,\lambda}$. More specifically, let us start with the \textit{ansatz}
\begin{equation}\label{eq: ansatz1}
f_1^{k,\lambda}=e^{iF_2^{k, \lambda}}\sum^\infty_{n=1}\frac{(-i)^{n+1}\Gamma_n}{\omega_k ^n},\qquad
f_2^{k,\lambda}=e^{iF_2^{k,\lambda}}\sum^\infty_{n=0}\frac{(-i)^n\tilde\Gamma_n}{\omega_k ^n},\qquad \Gamma_n,\tilde{\Gamma}_n\in\mathbb{R},
\end{equation}
where $\tilde{\Gamma}_0=1$ and the coefficients $\tilde\Gamma_n=\tilde\Gamma_n(\Gamma_1,...,\Gamma_{n-1})$ are fixed by the first condition in Eq. \eqref{eq:fg}. Then, for any $n\geq 0$, the non-diagonal part of the Hamiltonian is of order $\mathcal{O}(\omega_k^{-n-1})$ if  \cite{diagfermi}
\begin{align}\label{eq: recursive}
\Gamma_{n+1}=-\frac{Ma}{2}\tilde\Gamma_n+ \frac{a}{2}\sum_{l=1}^n\Big[\Gamma_l \{\tilde\Gamma_{n-l},H_{\vert 0}\}-\tilde\Gamma_{n-l}\{\Gamma_l,H_{\vert 0}\}-\frac{2}{a}\tilde\Gamma_l\Gamma_{n+1-l}-M(\Gamma_l \Gamma_{n-l}+\tilde\Gamma_l\tilde\Gamma_{n-l})\Big].
\end{align}
\end{itemize}
In all of these results, $\{ \boldsymbol{\cdot},\boldsymbol{\cdot}\}$ are the Poisson brackets of our truncated system and $H_{\vert 0}$ is the Hamiltonian constraint of the unpertubed FLRW cosmology. Recall also that $M$ is the bare mass of the Dirac field.

\section{Adiabatic fermionic states in the Weyl representation}\label{adiabatic}

In order to introduce the adiabatic scheme, we first limit our attention to situations in which the background variables are treated as classical functions of time which follow the Hamilton trajectories dictaded by $H_{\vert 0}$ (namely, by the Einstein equations in the linearized theory). In this way, we can express all of our fermionic variables in terms of a conformal time $\eta$, and work in the framework of QFT in a fixed FLRW cosmology. In addition, we restrict all considerations from now on to the trivial spin structure $\vec{\tau}=0$, as this is the choice that can be naturally extended to the case of non-compact spatial sections, which is precisely the scenario contemplated in Ref. \cite{barbero} for the construction of adiabatic states in the Dirac representation that we will parallel here, although now adopting the Weyl representation. Then, given a choice of initial time $\eta_0$, any set of annihilation and creation-like variables defined by Eqs. \eqref{eq: u} and \eqref{eq:fg} selects a decomposition of the Dirac field of the form
\begin{equation}\label{decomp}
\psi(\eta,\vec{x})
=
\sum_{\vec{k}\in\mathbb{Z}^{3}}\sum_{\lambda=\pm 1}
\left[u_{\vec{k},\lambda}(\eta,\vec{x})A_{\vec{k},\lambda} + v_{\vec{k},\lambda}(\eta,\vec{x})\bar{B}_{\vec{k},\lambda} \right],
\end{equation}
where we have defined the annihilation and creation-like constant coefficients
\begin{equation}\label{annicoef}
A_{\vec{k},\lambda}=a_{\vec{k},\lambda}(\eta_0), \qquad  \bar{B}_{\vec{k},\lambda}=\bar{b}_{-\vec{k},\lambda}(\eta_0),
\end{equation}
and
\begin{equation}\label{uv}
u_{\vec{k},\lambda}(\eta,\vec{x})=\frac{e^{i2\pi \vec{k}\vec{x}/l_0}}{\sqrt{ l_0^3 a^3}}\begin{pmatrix}
h_{k,\lambda}^{I}(\eta)\xi_{\lambda}(\vec{k}) \\ 
\lambda h_{k,\lambda}^{II}(\eta)\xi_{\lambda}(\vec{k})\end{pmatrix}, \qquad v_{\vec{k},\lambda}(\eta,\vec{x})=-e^{-iJ_{k,\lambda}(\eta_0)}\lambda\gamma^2 \bar{u}_{\vec{k},\lambda}(\eta,\vec{x}).
\end{equation}
The functions $(h_{k,\lambda}^{I},h_{k,\lambda}^{II})$ are a basis of mode solutions of the Dirac equation, and they are normalized so that $|h_{k,\lambda}^{I}|^2+|h_{k,\lambda}^{II}|^2=1$ (this normalization is just a consequence of the canonical anticommutation relations). Their explicit form in terms of the time-dependent coefficients that define the annihilation and creation-like variables in Eqs. \eqref{eq: u} and \eqref{eq:fg} is not needed yet, and hence we will postpone specifying it until the next section. We note that the spinors $v_{\vec{k},\lambda}$, that contain the information about antiparticles in the decomposition of the Dirac field are the charge conjugate of those that describe the particles, $u_{\vec{k},\lambda}$, only if we fix $J_{k,\lambda}(\eta_0)$ so that $v_{\vec{k},\lambda}=-\gamma^2\bar{u}_{\vec{k},\lambda}$. Although this is not necessary in principle, we choose to do so in order to maintain this charge conjugation symmetry in the selected Fock representation.

The identification of adiabatic states proposed  in Ref. \cite{barbero} for cosmological spacetimes was implemented in the Dirac representation of the Clifford algebra. Here we will instead obtain these states in the Weyl representation following the same line of reasoning, that we summarize below. Since the field $\psi$ is a solution to the Dirac equation, the variables $(h^{I}_{k,\lambda},h^{II}_{k,\lambda})$ in the decomposition \eqref{decomp}-\eqref{uv} satisfy the Schr\"odinger-like equation \cite{uniqfermi}
\begin{equation}\label{schr}
i\partial_\eta \boldsymbol{h}=\boldsymbol{H}(\eta)\boldsymbol{h},\qquad \boldsymbol{h}=\begin{pmatrix}h^{I}_{k,\lambda}\\h^{II}_{k,\lambda}\end{pmatrix}, \qquad \boldsymbol{H}=\lambda\begin{pmatrix}-\omega_k&Ma\\ Ma&\omega_k\end{pmatrix}.
\end{equation}
The construction of adiabatic states starts by diagonalizing the time-dependent Schr\"odinger Hamiltonian $\boldsymbol{H}(\eta)$. For this, one performs an explicitly time-dependent change of variables by means of a unitary matrix $\boldsymbol{U}_0$, such that the new variables $\boldsymbol{h}_0=\boldsymbol{U}_0^{\dagger}\boldsymbol{h}$ satisfy a similar equation, but with a lower dominant asymptotic order in (inverse) powers of $\omega_k$ in the non-diagonal part. A valid choice is the unitary matrix that brings $\boldsymbol{H}$ into its diagonal form $\boldsymbol{D}_0$. In this way, one obtains
\begin{equation}
i\partial_\eta \boldsymbol{h}_{0}=\boldsymbol{H}_{0}\boldsymbol{h}_0,\qquad \boldsymbol{H}_{0}=\boldsymbol{D}_{0}-i\boldsymbol{U}^\dagger_{0}\partial_\eta \boldsymbol{U}_{0}.
\end{equation}
This process can be repeated iteratively. At each step one gets the following new variables and Hamiltonian:
\begin{equation}
\boldsymbol{h}_{j+1}=\boldsymbol{U}^\dagger_{j+1}\boldsymbol{h}_j,\qquad
\boldsymbol{H}_{j+1}=\boldsymbol{D}_{j+1}-i\boldsymbol{U}^\dagger_{j+1}\partial_\eta \boldsymbol{U}_{j+1}.
\end{equation} 
The diagonal matrix $\boldsymbol{D}_{j+1}$ and the unitary matrix $\boldsymbol{U}_{j+1}$ are found diagonalizing $\boldsymbol{H}_j$, and then
$i\partial_\eta \boldsymbol{h}_{j+1}=\boldsymbol{H}_{j+1}\boldsymbol{h}_{j+1}$. The important point for the adiabatic scheme is that the dominant asymptotic order in the non-diagonal part of $\boldsymbol{H}_{j}$ decreases at each iterative step, in the limit $\omega_k\rightarrow \infty$. Therefore, the approximation of $\boldsymbol{h}_n$ by a solution $\tilde{\boldsymbol{h}}_n$ to the diagonal dynamics dictated by $\boldsymbol{D}_{n}$ gets more and more accurate for large $\omega_k$ as we increase the order $n$ of our adiabatic iteration. A straightforward integration of the diagonal evolution gives
\begin{align}\label{adicond}
\tilde{\boldsymbol{h}}_n(\eta)=\tilde{\boldsymbol{U}}_n(\eta,\tilde\eta_0) \mathfrak{h}(\tilde\eta_0),\qquad 
\tilde{\boldsymbol{U}}_n={\rm diag}\left(\exp\left(-i\int_{\tilde\eta_0}^\eta\Omega_n\right),\exp\left(i\int_{\tilde\eta_0}^\eta\Omega_n\right)\right),\qquad\mathfrak{h}(\tilde\eta_0)=\begin{pmatrix}1\\0\end{pmatrix},
\end{align}
where $\tilde{\boldsymbol{U}}_n$ is a diagonal matrix  and $\pm\Omega_n$ are the diagonal elements of $\boldsymbol{D}_n$. This frequency $\Omega_n$ is manifestly positive in the asymptotic regime of infinitely large $\omega_k$. Besides, the initial condition $\mathfrak{h}(\tilde\eta_0)$ was motivated in Ref. \cite{barbero} in order to select positive frequencies. With this choice, an adiabatic Fock representation of order $n$ is characterized as follows by a specific basis of solutions $\boldsymbol{h}_{\vert n}(\eta)$ of Eq. \eqref{schr}, that we define in a similar way as for scalar fields \cite{luders}. They are determined precisely by the initial conditions at time $\eta_0$ obtained from the approximate solution at order $n$ after undoing all the changes of variables involved in the iterative process:
\begin{equation}\label{adinit}
\boldsymbol{h}_{\vert n}(\eta_0)=\left(\prod_{i=0}^n\boldsymbol{U}_i(\eta_0)\right)\tilde{\boldsymbol{U}}_n(\eta_0,\tilde\eta_0) \mathfrak{h}(\tilde\eta_0).
\end{equation}
Given the specific form of $\mathfrak{h}(\tilde\eta_0)$, different choices of initial time for the integration of the diagonal dynamics only yield different constant global phases in the expansion of the Dirac field $\psi$ in terms of annihilation and creation operators. Actually, these phases carry no relevant information about the quantum properties of the field, and so we can choose them freely and set $\eta_0={\tilde\eta}_0$ for simplicity.

In the above discussion, we have applied the adiabatic procedure directly to the decomposition \eqref{decomp}-\eqref{uv} of the fermionic field in the Weyl representation of the Clifford algebra. Let us now show that the result coincides indeed with that obtained in Ref. \cite{barbero} employing the same type of decomposition in the Dirac representation (and, therefore, starting with a different Schr\"odinger Hamiltonian). The change to the unitarily related Weyl representation can be carried out as follows:
\begin{equation}
T\gamma^{b}_{D}T^{\dagger}=\gamma^b_W,\qquad T=\frac{1}{\sqrt{2}}\begin{pmatrix}I &-I \\ I & I\end{pmatrix}.
\end{equation}
In the rest of this section, the sub/superscripts $D$ and $W$ indicate spinors in the Dirac or the Weyl representation, respectively. Thus, for the fermionic field, we have $\psi^W=T\psi^D$ or, in terms of the basis of mode solutions associated with a certain vacuum,
\begin{equation}
\boldsymbol{h}^W=\tilde{T}\boldsymbol{h}^D,\qquad \tilde{T}=\frac{1}{\sqrt{2}}\begin{pmatrix}1 &-\lambda \\ \lambda & 1\end{pmatrix}, 
\end{equation}
where $\tilde{T}$ is clearly unitary, because $\lambda^2=1$. The Schr\"odinger Hamiltonians in both representations are then unitarily related by $\boldsymbol{H}^W=\tilde{T}\boldsymbol{H}^{D}\tilde{T}^{\dagger}$, and therefore they have the same diagonal form $\boldsymbol{D}_0$. It follows that the zeroth-order step in the adiabatic iterative procedure is the same when applied to both representations, except for the unitary matrix that diagonalizes the Hamiltonian, which changes as $\boldsymbol{U}_0^{W}=\tilde{T}\boldsymbol{U}_0^{D}$. Since this transformation is unitary and constant, the Schr\"odinger Hamiltonian $\boldsymbol{H}_0$ needed for the next adiabatic step is already the same at zeroth-order, regardless of whether one applies the procedure in the Dirac or the Weyl representation. Hence, the same quantities must appear as well in both representations for all the higher-order steps up to the desired order $n$, including the conditions \eqref{adicond} on $\mathfrak{h}(\tilde\eta_0)$. It is then straightforward to conclude what we wanted to check, namely that, for an adiabatic state of order $n$, one obtains the same set of solutions independently of whether one first performs the adiabatic construction in the Dirac representation and then transforms to the Weyl representation, or alternatively  one applies the construction directly in the latter of these representations, with the change between them given by the transformation $\boldsymbol{h}_{|n}^{W}=\tilde{T}\boldsymbol{h}_{|n}^{D}$.

\section{Unitary equivalence and choice of initial time}\label{unit}

The family of Fock representations for the Dirac field presented in Sec. \ref{model} are completely characterized by certain background-dependent (or time-dependent, in the context of QFT in the linearized theory) functions $f_{1}^{k,\lambda},f_{2}^{k,\lambda}$, and $J_{k,\lambda}$, subject to the conditions \eqref{eq:fg}. In terms of them, the basis of mode solutions for the field decomposition \eqref{decomp}-\eqref{uv} adopts the expression
\begin{align}\label{hfg}
\boldsymbol{h}(\eta)=
\left[I-\frac{1-\lambda}{2}(I-i\sigma_{2})\right]\begin{pmatrix}\bar{f}_1^{k,\lambda}(\eta)\alpha_{k,\lambda} (\eta,\eta_0) - f_2^{k,\lambda}(\eta)e^{-iJ_{k,\lambda}(\eta_0)}\bar{\beta}_{k,\lambda} (\eta,\eta_0) \\ \bar{f}_2^{k,\lambda}(\eta)\alpha_{k,\lambda} (\eta,\eta_0) + f_1^{k,\lambda}(\eta)e^{-iJ_{k,\lambda}(\eta_0)}\bar{\beta}_{k,\lambda} (\eta,\eta_0)\end{pmatrix},
\end{align}
where we have taken into account that, for QFT in curved spacetimes, the evolution of the variables defined in Eqs. \eqref{eq: u} and \eqref{eq:fg} comes from the dynamics dictated by the Dirac equation, and is given by a Bogoliubov transformation of the form \cite{uniqueness3,uniqfermi}
\begin{align}
&a_{\vec{k},\lambda}(\eta)=\alpha_{k,\lambda} (\eta,\eta_0) a_{\vec{k},\lambda}(\eta_0)+\beta_{k,\lambda}(\eta,\eta_0)\bar{b}_{\vec{k},\lambda}(\eta_0) , \\ \nonumber 
&\bar{b}_{\vec{k},\lambda}(\eta)=e^{i[J_{k,\lambda}(\eta)-J_{k,\lambda}(\eta_0)]}\bar\alpha_{k,\lambda} (\eta,\eta_0) \bar{b}_{\vec{k},\lambda}(\eta_0)-e^{i[J_{k,\lambda}(\eta)-J_{k,\lambda}(\eta_0)]}\bar\beta_{k,\lambda}(\eta,\eta_0)a_{\vec{k},\lambda}(\eta_0),
\end{align}
with $|\alpha_{k,\lambda}|^2+|\beta_{k,\lambda}|^2=1$. From these relations, it is then clear that any adiabatic state defined by the initial conditions \eqref{adinit} for $\boldsymbol{h}_{\vert n}(\eta)$ at time $\eta_0$ (equal to $\tilde{\eta}_0$, for simplicity) is associated to a choice of functions $f_{1\vert n}^{k,\lambda}$ and $f_{2\vert n}^{k,\lambda}$ such that
\begin{align}\label{fad}
\begin{pmatrix}
\bar{f}_{1\vert n}^{k,\lambda}(\eta_0) \\ \bar{f}_{2\vert n}^{k,\lambda}(\eta_0)
\end{pmatrix}=\left[I-\frac{1-\lambda}{2}(I+i\sigma_{2})\right]\left(\prod_{i=0}^n\boldsymbol{U}_i(\eta_0)\right) \mathfrak{h}(\eta_0).
\end{align}
Here, we have used that $\alpha_{k,\lambda}(\eta_0,\eta_0)=1$ and $\beta_{k,\lambda}(\eta_0,\eta_0)=0$. The quantities $f_{1\vert n}^{k,\lambda}(\eta_0)$ and $f_{2\vert n}^{k,\lambda}(\eta_0)$ will depend, in general, on the scale factor of the homogeneous cosmological background and its derivatives, evaluated at time $\eta_0$.  Extending the dependence of these homogeneous variables on the initial time $\eta_0$ to the whole time domain will indeed define a set of annihilation and creation-like variables in the same way as in Eqs. \eqref{eq: u} and \eqref{eq:fg}, up to the choice of the time-dependent phases $J_{k,\lambda}$ and $F_{2}^{k,\lambda}$. Actually, it is worth noting that the initial value of these phases at time $\eta_0$ is already fixed, respectively, by imposing charge conjugation symmetry and by relation \eqref{fad}.

Let us now analyze the properties of the resulting adiabatic quantization and its associated annihilation and creation operators. With respect to the asymptotic expansion in the limit $\omega_k\rightarrow\infty$, the adiabatic construction is such that $f_{1\vert n}^{k,\lambda}$ mantains, for each $n\geq 1$, the same dominant terms that appear in $f_{1\vert n-1}^{k,\lambda}$ up to order $\mathcal{O}(\omega_k^{-n-1})$. Computing just the two first adiabatic orders, one observes that
\begin{align} \label{eq:adiab1}
f_{1\vert 0}^{k,\lambda}(\eta)&=\frac{Ma(\eta)}{2\omega_k}+\mathcal{O}(\omega_k^{-2}),\\
\label{eq:adiab2} f_{1\vert 1}^{k,\lambda}(\eta)&=\frac{Ma(\eta)}{2\omega_k}+\frac{iM  a'(\eta)}{4\omega_k^2}+ \mathcal{O}(\omega_k^{-3})=\frac{ M a(\eta)}{2\omega_k}-i\frac{\pi M\pi_a(\eta)}{3l_0^3\omega_k^2}+ \mathcal{O}(\omega_k^{-3}).
\end{align}In the last line we have denoted with a prime the total derivative with respect to the conformal time, and used Hamilton equations for the homogeneous cosmology in the linearized theory in order to express the result in terms of canonical variables. The dominant terms in these expressions (that are written explicitly) will remain in higher-order adiabatic states, according to our comments. Recalling then the results listed in Sec. \ref{model}, and in particular condition \eqref{eq: cond1}, we can see just from the zeroth-order shown in Eq. \eqref{eq:adiab1} that all the adiabatic states live in the  family of unitarily equivalent vacua that are determined by the annihilation and creation-like variables \eqref{eq: u} and \eqref{eq:fg},  for which the quantum Heisenberg evolution is unitarily implementable. Furthermore, for adiabaticity order greater than zero, the Fock quantization of these annihilation and creation-like variables leads to a finite mean backreaction in hybrid quantum cosmology (in the sense explained in Sec. \ref{model}) and their contribution to the total Hamiltonian constraint of the system is well defined on the dense set of Fock space spanned by the states with definite number of particles/antiparticles.

Finally, let us comment on the relevance of the choice of initial time $\eta_0$ in the discussed construction of fermionic adiabatic states. Indeed, each of such adiabatic representations of the Dirac field depends on the time at which one sets initial conditions of the form \eqref{adinit} for the basis of mode solutions. Let us specifically call $\boldsymbol{h}_{|n}^{\eta_0}$ the basis of adiabatic solutions obtained with initial conditions at time $\eta_0$. Imagine that, rather than at $\eta_0$, we imposed adiabatic initial conditions at another time $\eta_1$, getting in that way a new basis of mode solutions $\boldsymbol{h}_{|n}^{\eta_1}$. According to our discussion above [and in particular to formula \eqref{hfg}], the two sets of solutions, evaluated at the same time $\eta_0$, will be related by
\begin{align}
\boldsymbol{h}_{|n}^{\eta_1}(\eta_0)=\left[I-\frac{1-\lambda}{2}(I-i\sigma_{2})\right][\alpha_{k,\lambda}(\eta_0,\eta_1)\boldsymbol{h}_{|n}^{\eta_0}(\eta_0)-i\lambda\sigma_{2}\bar\beta_{k,\lambda}(\eta_0,\eta_1)\bar{\boldsymbol{h}}_{|n}^{\eta_0}(\eta_0)],
\end{align}
where we have fixed $J_{k,\lambda}(\eta_1)=(3+\lambda)\pi/2$ by requiring charge conjugation symmetry. This relation between the two sets of data at $\eta_0$ is a Bogoliubov transformation, and its unitary implementability in the quantum theory depends exclusively on the square summability of the beta coefficients, over all $\vec{k}\in\mathbb{Z}^3$. But we note that, in norm, these coefficients are precisely the same that characterize the dynamical transformations of the annihilation and creation-like variables, whose evolution that we have seen that indeed is unitarily implementable. Hence, we conclude that any two adiabatic representations that differ on the value of the initial time at which one imposes the conditions \eqref{adinit} are unitarily equivalent. Furthermore, this equivalence is directly related to the fact that the representations allow the definition of families of annihilation and creation operators that can evolve unitarily.

\section{Conclusions}\label{conclusions}

In this work, we have investigated the relation between the adiabatic construction and the criterion employed in hybrid quantum cosmology to select Fock states that can play the role of vacua for the Dirac field, treated as a fermionic perturbation of an inflationary flat FLRW universe. Specifically, we have found that all adiabatic states belong to the family of unitarily equivalent Fock vacua employed in hybrid quantum cosmology, characterized by the invariance under the isometries of the spatial sections and by a unitarily implementable Heisenberg evolution of the corresponding annihilation and creation operators when the FLRW cosmology is regarded as a curved background. Moreover, for adiabatic orders other than zero, they allow quantizations with other desirable ultraviolet properties, such as a finite backreaction term in the only non-trivial constraint of the system and a properly defined fermionic Hamiltonian operator.

Given a mode decomposition of a solution to the Dirac equation, its coefficients determine a set of annihilation and creation constant operators. The adiabatic scheme that we have discussed makes use of this fact, selecting a particular set of mode solutions. More specifically, any decomposition is characterized by functions that satisfy a Schr\"odinger-like equation with a time-dependent Hamiltonian matrix. One can introduce a series of time-dependent transformations on these functions that decrease the asymptotic order of the non-diagonal part of their Hamiltonian in the ultraviolet regime of large wave numbers. If one neglects this non-diagonal part once a certain asymptotic order is reached, it is straightforward to construct a set of approximate solutions and, in this way, specify a mode decomposition. In this work, we have adapted this procedure to the Weyl representation of the Clifford algebra. The implementation in the Dirac representation had been studied in Ref. \cite{barbero}. We have provided the transformation between these two representations and shown that the conclusions obtained in both cases are consistent.  

We have computed explicitly the approximate mode solutions at the two lowest adiabatic orders and, with them, we have identified the dominant and first subdominant asymptotic terms for large $\omega_k$ in the functions that define the corresponding dynamical sets of annihilation and creation-like variables. Comparing these asymptotic terms with those that are characteristic of the family of Fock quantizations admissible in hybrid quantum cosmology, we have proven that all adiabatic vacua belong indeed to this family and, furthermore, that for adiabatic orders other than zero, those vacua can be associated with annihilation and creation operators that lead to well-defined mean backreaction contribution and fermionic quantum Hamiltonian in the only non-trivial constraint of the system. These results also ensure that the alternative adiabatic vacua constructed with different choices of initial time for the integration of the approximate mode solutions are all unitarily related.

In spite of the proven unitary equivalence between the two considered quantization schemes, it is worth commenting that the approach followed in hybrid quantum cosmology possesses a useful feature that in principle is missing in the adiabatic proposal. Indeed, in the former approach one starts by characterizing the set of admissible annihilation and creation-like variables, including their dynamical behavior, and therefore the genuine quantum fermionic excitations that have desirable physical properties. On the other hand, the adiabatic approach only defines a Fock representation of the Dirac field in terms of constant annihilation and creation operators. Without further information, there is no unambiguous way of isolating, from the evolution of the field, a Heisenberg dynamics with nice quantum behavior that dictates exclusively the dynamical transformations of those fermionic operators, separating them from the background dependence. Clearly, after one has introduced a dynamical family of annihilation and creation-like variables in the hybrid approach, one can also make the corresponding identification of adiabatic states. This advantage of the hybrid strategy in specifying quantum excitations of the field that are dynamically well behaved can be a potential help to understand the origin of the plausibly good ultraviolet properties of adiabatic states. In fact, we have already seen here that the unitarity of the Heisenberg dynamics of the fermionic operators in the hybrid approach is capable to explain the equivalence (up to unitary transformations) of all the adiabatic states, irrespectively of the time selected to set their initial conditions.

\appendix

\section{Some comments about the adiabatic scheme proposed by Hollands}\label{Hollands}

An alternative construction of adiabatic states has been proposed by Hollands in Ref. \cite{hollands}. The first step in this procedure is to find a pseudo-differential operator (see e.g. Refs. \cite{pdo,pdo2}) $T$ that factorizes the spinorial Klein-Gordon operator, namely
\begin{equation}\label{eq: dirac1}
-(in^\mu\nabla_\mu+iK+H)(in^\mu\nabla_\mu-H)=-(in^\mu\nabla_\mu+iK+T)(in^\mu\nabla_\mu-T) 
\end{equation}  modulo an operator with smooth kernel. In this relation, \textit{H} is the one-particle Dirac Hamiltonian, $K=\nabla^\nu n_\nu$ is the extrinsic curvature of the spatial sections, and the operator \textit{T} has principal symbol $\sigma_1(T)(\vec x, \vec \xi)=\sqrt{h_{ij}(\vec x) \xi^i \xi^j}$, where $h_{ij}$ is the metric of the spatial sections. Although finding $T$ is a hard problem in general, one can construct approximate solutions by means of an iterative method. We call $T_n$ the resulting operator after \textit{n} steps. One then defines $L_{n,\pm}=T_n \pm H$ and looks for a positive hermitic operator $Q_n$ such that
\begin{equation}\label{eq:construction2}
L_{n,+}Q_nL_{n,+}^*+L_{n,-}^*Q_nL_{n,-}=1. 
\end{equation} 
With this, one can define the following operators:
\begin{equation} \label{eq:construction3}
B_n= L_{n,+}Q_nL_{n,+}^*,\qquad 
B_{n,-}=L_{n,-}^*Q_nL_{n,-},
\end{equation} 
which must be symmetric and positive. These operators determine the algebraic state desired for the quantization of the Dirac field \cite{hollands}. In fact, such a state corresponds to a Fock representation if and only if $B_n$ is a projector \cite{hollandspr}. In practice, to find these operators, it is convenient to introduce their mode decomposition. This was done in Ref. \cite{hollands} by using the Dirac representation of the Clifford algebra and a basis of spinors for which the one-particle Hamiltonian is instantaneously diagonal:
\begin{gather}\label{eq: base2}
u_{\vec k,\lambda}^+=\frac{\mathcal{U}_{ k, \lambda}}{\sqrt{l_0^3 a^3}} \begin{pmatrix} \xi_{\lambda} (\vec{k}) \\ 0
\end{pmatrix}e^{i2\pi(\vec k +\vec \tau)\vec{x}/l_0},\qquad u_{ k,\lambda}^-=\frac{\mathcal{U}_{ k, \lambda}}{\sqrt{l_0^3 a^3}} \begin{pmatrix} 0 \\ \xi_{\lambda} (\vec{k})
\end{pmatrix}e^{i2\pi(\vec k +\vec \tau)\vec{x}/l_0}, 
\end{gather}
where, defining $\Delta_k(a)=\sqrt{\omega_k^2+M^2 a^2}$, we have called
\begin{gather}
\mathcal{U}_{k, s}=\frac{1}{\sqrt{2\Delta_k(a)}}\begin{pmatrix} \sqrt{\Delta_k(a)+Ma} & -\lambda \sqrt{\Delta_k(a)-Ma}  \\ \lambda\sqrt{\Delta_k(a)-Ma} & \sqrt{\Delta_k(a)+Ma}
\end{pmatrix}.
\end{gather} 
With this basis one may define the mode decomposition of any differential operator $B$ on the spatial sections by the formulas
\begin{equation}\label{eq: descomposicionbase2}
a^3\int_{T^3} d^{3}\vec{x}\, f_1^{\dagger}B f_{2} = \sum_{\vec{k},s,pq}  b_{\vec k, s}^{pq} \bar{\tilde f}^p_{1,\vec k, \lambda} \tilde f^q_{2,\vec k, \lambda},\qquad \tilde{f}^p_{\vec{k},\lambda}=a^3\int_{T^3} d^{3}\vec{x} \,(u_{\vec k,\lambda}^p)^{\dagger}f,
\end{equation}
for any two spinors $f_1$ and $f_2$, with $p,q=\pm$. In essence, this decomposition maps operators (and pseudo-differential operators) into 2x2 complex matrices while respecting products and the adjoint operation. 

In the following, to simplify our notation, we will drop from it the dependence on $\vec k$ and $\lambda$ unless explicitly stated. In addition, we will use lowercase letters to refer to the mode decomposition of the operators, with the correspondence $T_n\rightarrow \tau_n$, $H\rightarrow h$, $Q_n \rightarrow q_n$, $L_n\rightarrow \ell_n$, and $B_n \rightarrow b_n$. Besides, we recall that the prime symbol denotes the total derivative with respect to the conformal time. The decomposition \eqref{eq: dirac1}, as given by Eq. \eqref{eq: descomposicionbase2}, can then be re-expressed as
\begin{equation}\label{eq: recurrence}
i { \tau'} + \frac{3ia'}{2a} \tau+ [\tau, d]+ a \tau^2=i {h'} + \frac{3ia'}{2a} h+ [h, d]+a h^2,
\end{equation} 
where
\begin{equation}
d=i\ \mathcal{U}^*(\partial_{\eta}\ \mathcal{U})= \frac{\lambda\omega_k M   a'}{2 (\omega_k^2+ M^2 a^2)}\sigma_2.
\end{equation} 
The procedure to determine $\tau_n$ goes as follows. Starting from the \textit{ansatz} $ \tau_n=\sum_{j=0}^n  \vartheta_j,$ with $  \vartheta_j=\mathcal{O}(\omega_k^{1-j})$, and setting $ \tau_0=\text{diag}\ [\sqrt{a^{-2}\omega_k^2+ M^2 },\sqrt{a^{-2}\omega_k^2+M^2 }]$, one solves \eqref{eq: recurrence} iteratively, obtaining
\begin{equation}\label{eq: recursapp}
\vartheta_{n+1}=\frac{1}{2\sqrt{\omega_k^2+M^2 a^2}}\left[F(h)-F(\tau_n)\right], 
\end{equation} 
where we have defined $F(o)= i  o' + 3i (\ln{a})' o /2+ [o,d]+a o^2$. The mode versions of Eqs. \eqref{eq:construction2} and \eqref{eq:construction3} are then used to construct $b_n$. To be able to compare the algebraic state resulting from the operator $B_n$ with our family of unitarily equivalent Fock vacua, that operator must be a non-trivial projector, something that requires that $b_n$ be singular. Unfortunately this turns out not to be the case in the system that we are considering, as can be checked by noticing that, for all $n\geq 1$,
\begin{eqnarray}
\ell_{n,+}=\text{diag}\left[\frac{2\omega_k}{a},-\frac{ia'}{2a}\right]+\mathcal{O}(\omega_k^{-1}),\qquad q_{n}=\text{diag}\left[\frac{a^2}{4\omega_k^2},\frac{a^2}{4\omega_k^2}\right]+\mathcal{O}(\omega_k^{-3}) .
\end{eqnarray}  
Clearly, this result implies that $\det(b_n)\neq 0$ $\forall n\geq 1$, except for the trivial case of a constant scale factor, therefore contradicting the assumption that it is singular.

\acknowledgements

This work was supported by Project. No. MINECO FIS2017-86497-C2-2-P from Spain.

\end{document}